\documentclass{article}
\usepackage{graphicx} % Required for inserting images
\usepackage{float}
\usepackage{multicol}
\usepackage[a4paper, total={7in, 10in}]{geometry}
\usepackage{listings}
\usepackage{xspace}
\usepackage{url}
\usepackage{cleveref}

\title{\proto: Detection and prevention of Byzantine behaviour in DAG-based consensus protocols}
\author{Andrey Chursin}
\date{Aug 2024}

\begin{document}
\newcommand{\proto}{\textsc{Adelie}\xspace}

\maketitle
\begin{multicols*}{2}

\begin{abstract}
Recent developments in the Byzantine Fault Tolerant consensus protocols have shown the DAG-based protocols to be a very promising technique. While early implementations of DAG-based protocols such as Narwhal/Bullshark\cite{narwhal}\cite{bullshark}\cite{shoal}\cite{hammerhead} trade high throughput for a low latency, the latest versions of DAG-based protocols such as Mysticeti\cite{mysticeti} and Shoal++\cite{shoalpp} show that indeed a latency comparable to that of traditional consensus protocols such as HotStuff\cite{hotstuff}\cite{monad}\cite{pbft} can be achieve with the DAG-based consensus protocols while still maintaining high throughput. Mysticeti in particular achieves a low latency by implementing a novel approach of using an uncertified DAG - a significant breakthrough comparing to the certified DAG used in the previous generations of the protocol. However, the uncertified DAG exposes the system to new vectors of attacks by Byzantine validators that did not exist in the certified DAG protocols.
\par In this paper we describe those issues and present the \proto protocol, that addresses issues that comes with an uncertified DAG. We also incorporate some of the techniques from the Shoal++ to reduce latency even further. This paper also presents an implementation of \proto protocol - \textbf{bftd} that demonstrates yet another breakthrough in the maximum achieved TPS and low latency.
\end{abstract}
\section{Introduction}
\par This paper presents \proto, a DAG-based Byzantine Fault Tolerant consensus protocol that achieves unparalleled performance combined with low latency. \proto protocol works on top of uncertified DAG, building on top of Mysticeti consensus protocol(specifically Mysticeti-C variation). Leader based DAG consensus protocols achieve consensus by denoting some blocks of the DAG as leader blocks and deriving sequence of such leaders by applying a so called "commit rule" on the DAG. By employing novel uncertified DAG approach, Mysticeti achieves leader commit latency of 1.5 RTT, on par with classical BFT consensus protocols such as HotStuff. Using an uncertified DAG however comes with the risk of exposing the protocol to various denial-of-service attacks by malicious validators. In this paper, we will go over the attack vectors made possible due to use of uncertified DAG, and present a novel way to prevent such attacks. Furthermore, \proto is one of the few BFT protocols that performs an active detection of a Byzantine behavior and allow correct validators to detect and cut off validators that have equivocated.

\par \proto protocol is an evolution of the Mysticeti protocol - we employ same basic DAG structure as Mysticeti protocol but introduce additional restrictions on top of Mysticeti DAG. In other words, any valid \proto DAG is also a valid Mysticeti DAG (but not the other way around). As such, \proto can use proven commit rules of the Mysticeti protocol. Hence, we do not discuss safety and commit rules in this paper and refer to safety proofs of the Mysticeti protocol. The bftd\footnote{https://github.com/andll/bftd} implementation uses the Mysticeti commit rule, optimally configuring it for a low latency, by implementing the approach proposed in the Shoal++ protocol, where all validators are denoted as leaders, which is shown to reduce the latency by up to 0.5 RTT.
\par We show in the paper that additional restrictions provided by the \proto protocol address several significant liveness issues of the Mysticeti protocol but do not hurt performance.
\section{Uncertified DAG}
\par Uncertified DAG is a novel approach that is proven to significantly reduce the commit latency by somewhat increasing complexity of the commit rule on the DAG.
\par Early generation DAG based protocols (such as Narwhal/Bullshark) use the so called \textit{certified} DAG. Each proposed block is first signed by a quorum of nodes, before it can be included in the DAG. Certification simplifies consensus protocol as it ensures only one block per \textsc{(Validator, Round)} ever certified. However, certification process comes with significant latency cost, as it requires 1.5 RTT to produce a single DAG element and at least 2 DAG elements are needed to perform a commit.
\par In contrast, Mysticeti protocol does not need a block to be certified. It only takes 0.5 RTT to produce an element of Mysticeti DAG. However, this means that validator can produce and include in the DAG multiple blocks at the same \textsc{(Validator, Round)}. Mysticeti protocol addresses safety implications of uncertified DAG with it's commit rule - even when malicious validator proposes equivocating blocks, the commit rule can still consistently order them.
\par However, there are new attacks that can be exploited
by the Byzantine validators that come with the use of uncertified DAG: \textbf{Flooding the DAG} and \textbf{the Phantom DAG}. We will discuss those attacks in detail below.
\subsection{Flooding the DAG}
    Validator can repeatedly equivocate, producing at least NumberOfValidators blocks on each round by sharing a different block with each validator. All of those blocks would have to be included in the DAG, because correct validators don't get a chance to compare what they received from others until the next round. As a consequence, malicious validators get an unfair advantage(by including significantly more transactions then correct validators) and can produce an outsized load on the system. Because we assume up to \textit{f} validators to be malicious and each can generate up to \textit{3f+1} blocks per round, the Byzantine validators alone can generate ~\textit{\(3f^2+f\)} blocks on each round, compared to \textit{2f+1} blocks collectively generated by all correct validators. \par To put this into perspective, on the cluster of 100 validators, correct validators would collectively generate \textbf{67 blocks per round}, whereas byzantine validators could collectively generate up to \textbf{3267 blocks per round}(33 validators sending 1 different block for each of 99 validators).
\subsection{Phantom DAG}
   Validators in the Mysticeti protocol can only generate next element of the DAG when they hear from \textit{2f+1} validators from a previous round. However, there is no mechanism that would force them to share the generated block. Byzantine validator can hold on to it's generated blocks, creating an entire sub-DAG without sharing. We will call such sub-DAG a phantom DAG. They can then choose to share the entire phantom DAG at some point, forcing correct validators to include an unbounded amount of blocks from a single source. What is worse, validators don't need to be intentionally malicious to perform such phantom DAG attack. Validators can manifest such behavior due to a simple network misconfiguration - when they can hear from other validators but can't share their generated blocks. The danger of including unbounded phantom DAGs from Byzantine validators does not only increase the load on the state synchronization and network, but the effects are also propagated further - many of such blocks would have to be committed in a single commit creating more overhead for the commit rule and the application. This can cause further denial of service cascade if some of the downstream logic cannot handle an oversized commit.
   \par This attack might be further complicated when Byzantine validators cooperate with different roles - one validator can generate large phantom DAGs and others can quickly include those phantom DAGs in their otherwise correct blocks, making it harder to implement basic rate-limiting strategies to protect against this behaviour.

\section{\proto protocol}
\subsection{General assumptions}
\par \proto protocol operates on a pre-defined set of validators. Each validator has a certain positive integer associated with it called \textit{stake}. The protocol operates correctly if the total stake of malicious validators does not exceed some value \textit{f}, and the total stake of all validators is at least \textit{3f + 1}.
\par Validators in \proto protocol generate blocks, sign them and share those blocks with other validators. Validators can \textit{include} other blocks as parents by referencing them in the \textit{parents} field of the generated block.
\par Each block contains certain metadata (described below) and payload opaque to the protocol. The payload would typically contain list of transactions interpreted by the higher level blockchain protocol.
\par The commit rule of the protocol ensures that at some point each correct validator can derive the same sequence of blocks \{X1, X2, ..., Xn\}, called leader sequence.
\par Validators can then use the produced leader sequence to linearize the DAG and derive the same total order across committed blocks.
\par If validators interpret payload of the block as a list of transactions, they can then derive a consistent total order of the transactions in the given DAG(in the committed part of it).
\subsection{Block structure}
\par Validator only accept blocks from another validator if the block is valid according to protocol rules, and if validator knows all the blocks if the sub-DAG of a given block.
\par Each block contains the following data:
\begin{itemize}
\item \textsc{Round} - a non-negative integer
\item \textsc{Author} - identifier of the validator
\item \textsc{parents} - list of references of parent blocks
\item \textsc{Payload}
\item \textsc{Signature} - an Ed25519 signature produced by the Author, over the data above
\item \textsc{Hash} - Blake2 hash over data above
\end{itemize}
\par Block is identified by it's \textsc{reference} - the tuple of \textsc{(Round, Author, Hash)}.
\par The block with round 0 is called a \textit{genesis block}.
\subsection{Block validity rules}
\par The common set of rules for accepting the block in both \proto and Mysticeti protocols are the following:
\begin{itemize}
\item Each parent of the block has round lower then block's round.
\item A valid non-genesis block has a parent produced by the same author.
\item For a valid non-genesis block, the total stake of all parent blocks with round block.Round-1 should be equal or above quorum stake(2f+1). This is so called \textit{threshold clock rule}.
\item Correct validator should only produce up to one block per round.
\item When correct validator generates a new block, it always includes previously generated block as the new block's parent (in other words, correct validators always build the DAG on top of their previous blocks).
\end{itemize}
\par In addition to the rules above, \proto protocol requires that only up to one parent per validator can be included by the block.
\par To protect from the attacks described above, we also introduce two additional restrictions on the proposed blocks - the \textbf{block view rule} and a \textbf{critical block view}.
\subsection{Block view rule}
\par Definitions:
\par Block A is a \textbf{parent} of block B if block B lists reference(A) in the list B.parents
\par Block A is \textbf{referenced} by block B if block A is in the sub-tree of block B.
\par Block A is \textbf{preceding} block B if A is a parent of B and author(A) == author(B).
\par The \textbf{mainline} of block A is set of blocks formed as \{A, preceding(A), preceding(preceding(A)), ...\}. In other words, mainline of block A is a block A itself and all the preceding blocks.
\par The \textbf{block\_view(A)} for block A maps each validator V to some block block\_view(G)[V]:
\begin{itemize}
    \item For each genesis block G and validator V, block\_view(G)[V] := GenesisBlock(V)
    \item For each non-genesis block A,
\begin{lstlisting}[language=Python]
def block_view(A) :=
 v = block_view(preceding(A))
 for p in A.parents:
  v = merge_view(v, block_view(p))
  v[p.author] = merge_element(
    v[p.author], 
    p
  )
 return v

def merge_element(left, right):
 if left is None || right is None:
    return None
 if !same_mainline(left, right):
    return None
 if left.round > right.round:
    return left
 else
    return right

def same_mainline(left, rigth):
  if left.round > right.round:
    return 
     mainline(left).contains(right)
  else:
    return 
     mainline(right).contains(left)
    
def merge_block_view(a, b):
 c = map();
 for v in Validators():
  c[v] = 
   merge_element(
    block_view(a)[v],
    block_view(b)[v]
  )
 return c
\end{lstlisting}
\end{itemize}
\begin{figure}[H]
    \centering
    \includegraphics[width=1\linewidth]{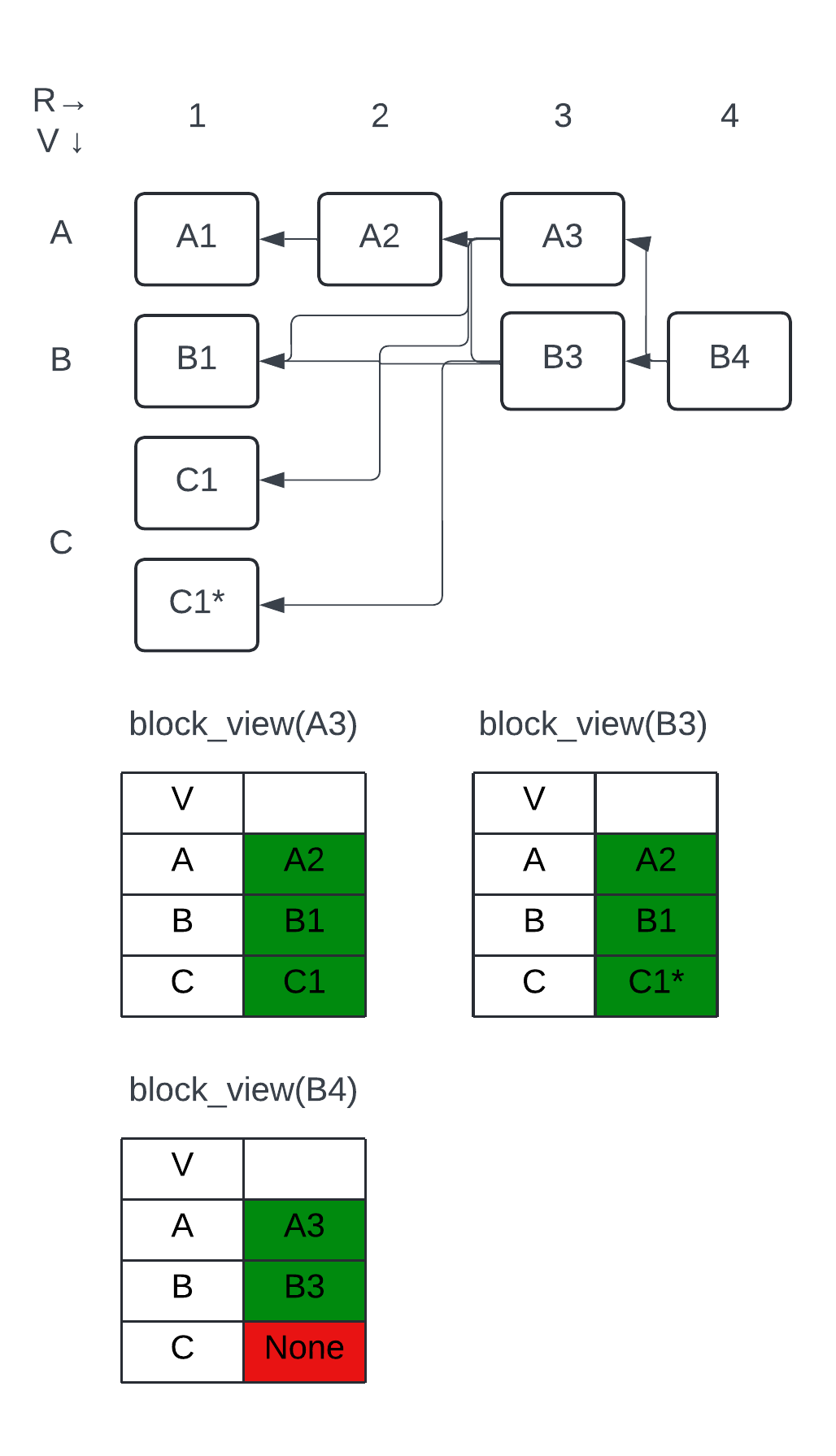}
    \caption{Block view detecting validator producing two different blocks at the same round}
    \label{Fig1}
\end{figure}
\begin{figure}[H]
    \centering
    \includegraphics[width=1\linewidth]{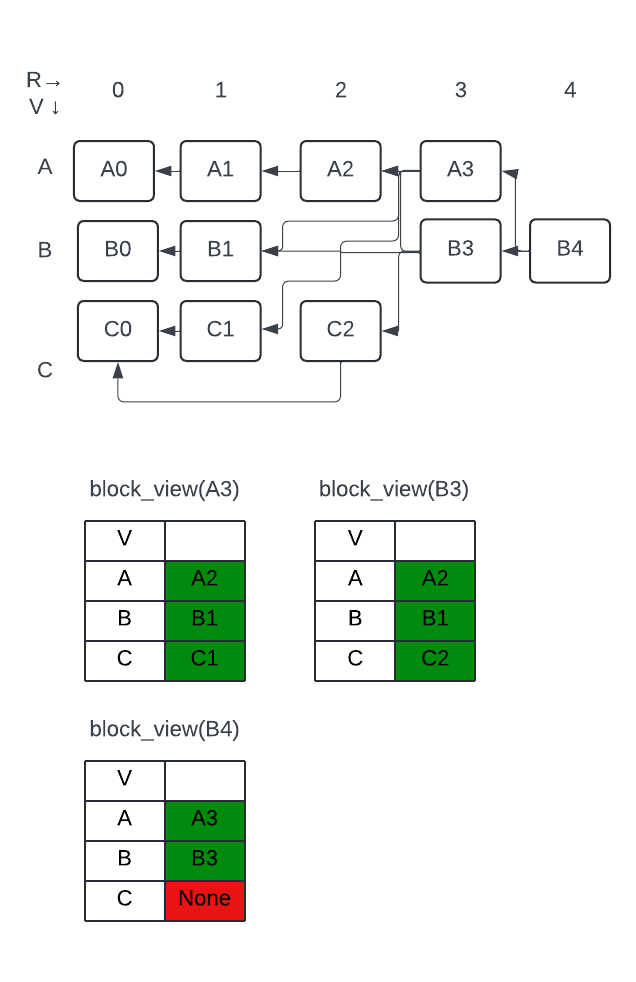}
    \caption{Block view detecting validator producing non-linear sequence of own blocks}
    \label{Fig2}
\end{figure}
\par It is worth pointing out few properties of the block\_view(A):
\begin{itemize}
    \item block\_view is not transmitted over network, instead it can be evaluated from the sub-DAG of a given block.
    \item Because block\_view(A) is only a function of A and its sub-DAG, all correct validator will evaluate same block\_view(A) for any block A.
    \item block\_view(A) can be cached when A is persisted locally, to avoid expensive computations.
\end{itemize}
\par \textbf{Theorem 1.} For any given block A, block\_view(A)[V] is None \textbf{if and only if} sub-DAG of A contains two blocks produced by V that are not part of the same mainline.
    \par (1) If all blocks produced by validator V in sub-DAG of A are part of one mainline, merge\_element will never return None, and therefore block\_view(A)[V] will not be None.
    \par (2) Let's say sub-DAG of A contains two blocks B and B` that are both produced by V, but are not part of the same mainline. Because sub-DAG of A contains B and B`, it also contains some blocks X with parents T and T`, such as sub-DAG of T only contains B and sub-DAG of T` only contains B`. block\_view(T)[V] will point to some block C, produced by validator V, and block\_view(T`)[V] will point to some block C`, also produced by V. Block B will be on mainline of C, and B` on mainline of C`. Because X references both T and T` as parents, merging block\_view(T) and block\_view(T`) on validator V will produce None, and block\_view(X)[V] will be set to None.
    Because block\_view(X)[V] is None, all blocks that contain X in the sub-DAG will have block\_view[V] set to None.
   Because sub-DAG of A contains X, block\_view(A)[V] is also None.
\par \textbf{Theorem 2.} For any block A and any correct validator V, block\_view(A)[V] is not None. (Even if A itself is produced by Byzantine validator):
\par (1) Because any two blocks produced by a correct validator are connected via mainline, block\_view(A)[V] is not None.
\par In other words, block\_view(A) maps validator V to its last produced block (as seen by A's sub-DAG) if V did not equivocate in A's sub-DAG, and None otherwise.
\par We can now formulate the \textbf{Block view rule}:
\begin{itemize}
    \item Block A can only include block B as a parent block if block\_view(preceding(A))[B.author] is not None.
\end{itemize}
\par In other words, block view rule does not allow to include blocks from validators that already have a provable equivocation as part of the known DAG.
\par It is worth pointing out that \textbf{block view rule} not only allows correct validators to quickly identify and block Byzantine validators, it enforces that no validator can help the Byzantine validator to disseminate it's blocks after the first equivocation.
\par Because of \textbf{Theorem 2}, the block view of a correct validator will never be set to None, meaning malicious validators can not manipulate the DAG(by arbitrary including correct validator's blocks) to force correct validator to appear to be malicious.

\subsection{Critical block rule}
\par The \textbf{critical block} for block A is defined as following:
\begin{itemize}
\item Let P be a preceding block for A. \item If P.round \textless A.round - 1, then P is critical block for A
\item If P.round == A.round - 1, then block preceding to P is a critical block for A.
\item Note, that some blocks such as genesis block and block immediately after genesis block do not have a critical block.
\end{itemize}
\begin{figure}[H]
    \centering
    \includegraphics[width=1\linewidth]{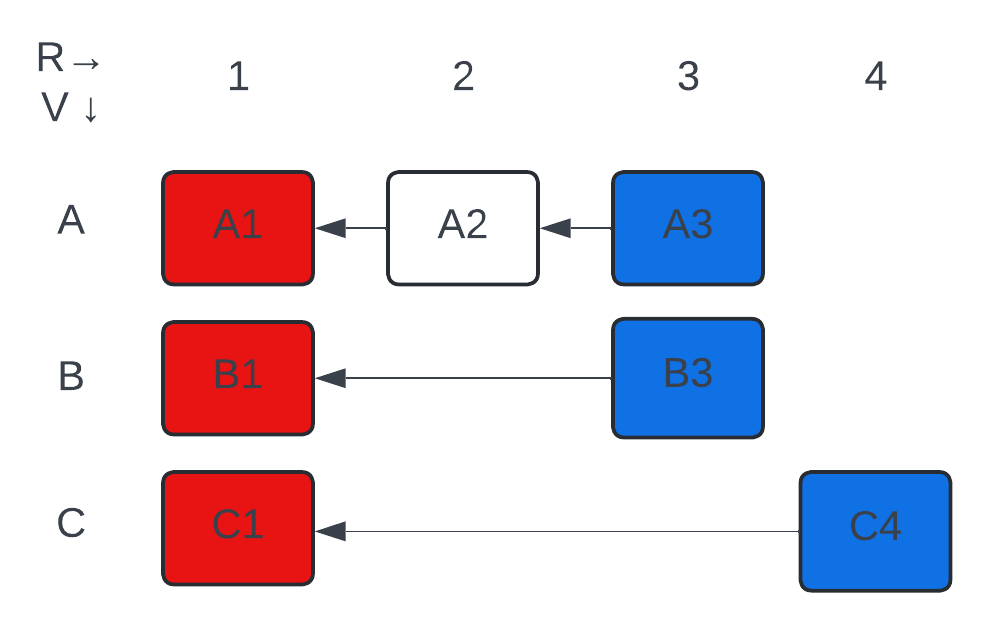}
    \caption{Each red block is a critical block for a corresponding blue block}
    \label{Fig3}
\end{figure}
The \textbf{critical block support} for block A is defined as following:
\begin{lstlisting}[language=Python]
def critical_block_support(A):
 v = author(A)
 C = critical_block(A)
 stake = 0
 for P in parents(A):
  w = block_view(P)
  if w[v] is None:
    return 0
  if w[v].round > C.round:
   stake += P.author.stake
 return stake
 \end{lstlisting}
\par We can then formulate \textbf{critical block rule} as following:
\begin{itemize}
    \item If block A has a critical block, the block A is only valid if \(critical\_block\_support(A) \geq validity\_stake\), where validity\_stake is f+1 stake.
\end{itemize}
\par Note that \textbf{correct validator(A) can always make progress in the system, if it can communicate with other correct validators}.
\begin{itemize}
    \item (1) This is because other correct validators will eventually include the last critical block of A, allowing it to produce a new block.
\end{itemize}
\par It is also worth pointing out that in a system that operates normally, the critical block rule does not slow down blocks creation as shown in the Evaluation section below.
\par For illustration purposes, we can think of the \textit{critical block rule} as similar to the \textit{threshold clock rule} from the Mysticeti protocol, but working in opposite direction. The \textit{threshold clock rule} requires validators to received some communication from the peers before it can produce a block on a new round. The \textit{critical block rule} requires validator to get their blocks included by others before they can produce more blocks.
\par Note that when \textit{block view rule} and \textit{critical block rule} are combined, each validator can only do up to O(number of validators) equivocations for the entire duration of an epoch. This is because with the \textit{critical block rule} Byzantine validator can only produce two rounds of blocks before sharing them with others, and with the \textit{block view rule} Byzantine validator's block view will be invalidated when equivocation is included in the shared DAG.
\subsection{Unlock critical block support}
While critical block rule normally does not affect system performance, it is theoretically possible under some network conditions for Byzantine validators to stall the network by advancing threshold clock beyond that of f+1 correct validators and becoming unresponsive after that. While this attack is hard to achieve in practice, we still need to design a way to unlock the protocol if this happens.
\par We achieve this by introducing a new message - \textit{Unlock(Validator, Round)}. When validator A reaches a certain timeout and cannot generate a new block, it generates and signs the Unlock message for each Validator and latest round of that validator they have seen. Validator A then sends signed \textit{Unlock(Validator, Round)} message to each corresponding validator.
\par Upon receiving f+1 unique Unlock messages for it's current round, validator is allowed to bypass the critical block support rule and generate a new block, by including the proof of f+1 \textit{Unlock} messages for a preceding block.
\section{Implementation}
\par We present \textbf{bftd} - a production ready Open Source implementation of the \proto protocol. bftd can achieve an unmatched performance even for the DAG-based consensus protocol - with a small cluster of 12 nodes of an inexpensive AWS instance type c7gn.xlarge, distributed across 4 regions globally (us-west, us-east, eu-west and ap-northeast) \textbf{bftd can deliver 450K TPS with a 630 millisecond finality}. (With a transaction size of 512 bytes).
\par bftd implementation uses \proto protocol to build the DAG and then applies the Mysticeti consensus rule to produce commits.
\par We configure the commit rule to use every validator as a leader on each round. As recently shown in Shoal++ paper\cite{shoalpp}, this can reduce the commit latency up to 0.5 RTT by being able to commit all blocks of the round at the same time.
\par Using every validator as a leader is also beneficial when coupled with the critical block rule, as it makes sure that critical block rule does not slow down leaders in their ability to generate new blocks.
\par Using \proto protocol for bftd not only helps with preventing denial of service attacks from malicious validators, it also largely simplifies catch up mechanisms (state synchronization) for the validators that lagged behind, as we don't need to deal with potential pathological DAG shapes like the phantom DAG described above.
\subsection{Architecture}
\par bftd achieves it's performance by employing number of proven techniques essential for high throughput and low latency:
\begin{itemize}
    \item bftd uses \textsc{RocksDB} to store blocks information and persists block views for each block added to the DAG.
    \item we efficiently cache the data and integrate the cache with the commit rule, ensuring a near 100\% hit rate for expensive operations such as running a commit rule or calculating the block view.
    \item we avoid serialization and de-serialization as much as possible, for example transactions in the block are not de-serialized unless needed by the application. This is useful because the majority of operations in the consensus core only need the block metadata and do not care about the payload.
    \item bftd uses TCP sockets encrypted with the noise protocol using \textsc{snow} library. We choose \textsc{ring} implementation for the noise cipher as it shows significantly better performance than the default rust implementation for the ciphers.
    \item the underlining network protocol used by bftd is inspired by the dist protocol in Erlang where nodes(validators) establish a full-mesh connections to each other and can send each other messages. The simple RPC mechanism is then built on top of such layer and is used for the catch up / state synchronization.
    \item using TCP sockets allows for a low overhead communication and enables the subscription model, where validators subscribe to each other blocks and when new block is created it is immediately pushed to everyone in the network. This approach plays a significant role in achieving low latency, comparing to the RPC push approach utilized in early implementation of narwhal/bullshark. 
    \item bftd does not use worker dissemination architecture that narwhal/bullshark has used to achieve scalability - such architecture is proven to over-complicate the system and can increase the latency if not implemented correctly. On the other hand, it is evident at this point that extremely high TPS can be achieved without a need for worker scaling.
    \item on the cryptography side, we use Ed25119 to sign blocks and Blake2b hash to calculate hash of the block. We also pre-hash block content with the Blake2b hash before signing it(as opposite to signing block content directly). It has been observed that SHA512 hash internally used by Ed25119 is very expensive when applied to large blocks, and hashing data with Blake2b before signing has shown to reduce CPU usage and latency by a significant amount.
    \item bftd uses combination of a single-thread \textit{core} component that operates on DAG and executes commit rule and an epoll-based asynchronous IO for networking. The single thread model was chosen for the core component because it makes testing significantly easier and most operations in the core do not benefit from parallelism.
    \item the \textit{syncer} component is responsible for various timeouts and network synchronization and is using \textit{tokio} runtime to perform async IO.
    \item expensive operations, such as verifying incoming blocks are also done as a part of the tokio pool, to minimize the load on the single-thread core component.
    \item on the persistence side, we configure RocksDB to use syscalls to perform every write but do not issue fsync calls during proposal. The consistency is critical when validator issues a proposal - it is important that validator "remembers" proposals it has shared with others, otherwise such validator would be Byzantine. The configuration of RocksDB we use ensures that proposals are persisted in the OS buffers before they are shared, which ensures that if application exits at any time, the required consistency guarantees will be met, because OS will eventually flush the buffers to disk. However, if machine experience physical crash or a kernel panic, it is possible that latest proposal(s) will be shared but not persisted, causing validator to equivocate upon restart. While this is a risk, it however seem like a reasonable and widely accepted trade off to achieve reasonable performance with a practical liveness.
\end{itemize}

\section{Evaluation}

\par We evaluate an implementation of \proto protocol - bftd in a real world deployment, using 12 AWS machines of instance type \textit{c7gn.xlarge} evenly distributed across 4 regions: us-west, us-east, eu-west and ap-northeast. 
\par \textit{c7gn.xlarge} is a lower tier instance type of a new generation machines equipped with 4 ARM-based virtual CPUs and 8 GiB of memory. We chose to test on the ARM CPUs as they provide good cost/efficiency ratio and at this point vast majority of projects in Rust ecosystem can compile under ARM architecture. 
\par For storage, we use EBS volumes with additionally provisioned 500 Mbps of throughput for each volume. The additional throughput is needed because of the sheer volume of the transactions being exchanged, and because RocksDB itself provides certain overhead when writing to disk. This overhead can be dealt with by using the write-ahead-log approach instead of writing to RocksDB as shown in the Mysticeti paper, but we chose not to use write-ahead-log in bftd at this time.
\par To assess performance, we introduce consistent load of given TPS for 5 minutes and measure p25, p50 and p75 transaction commit latency for the system. For the failure case scenario only data from the operation nodes is taken into account.
\par We demonstrate the following performance with the 100\% nodes operational:

\begin{figure}[H]
    \includegraphics[width=\linewidth]{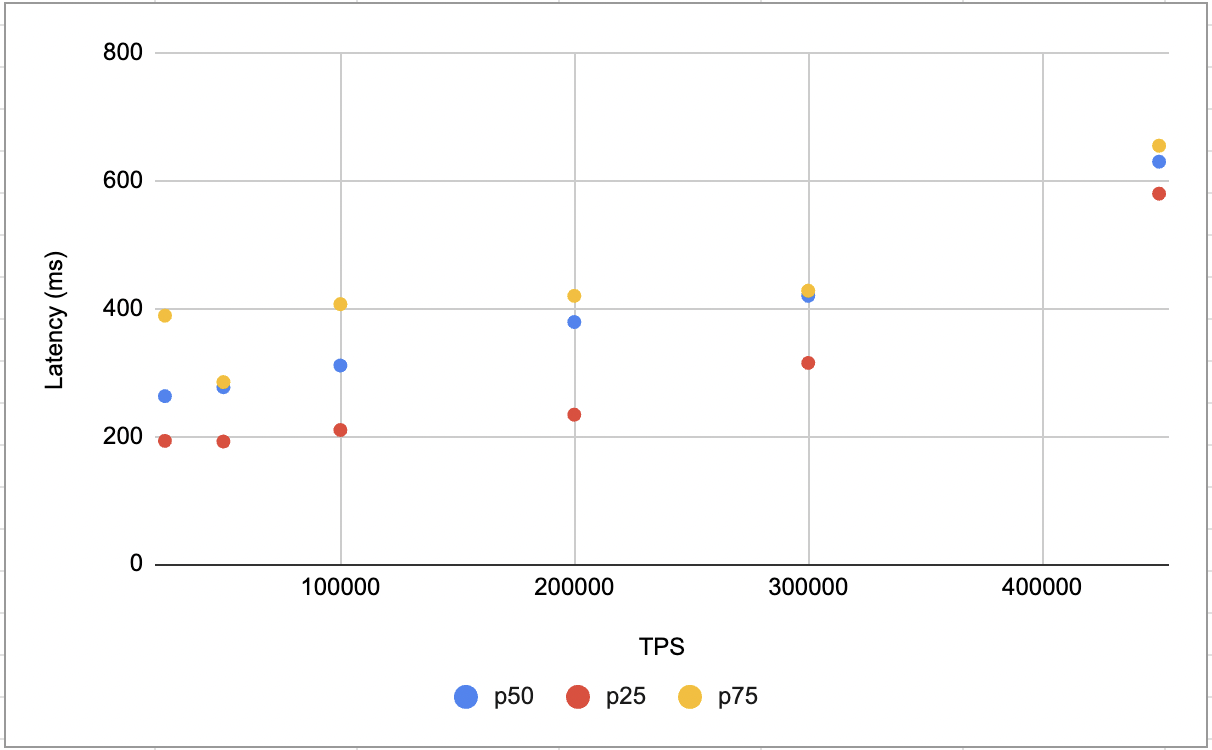}
    \caption{Latency with 100\% validators up}
    \label{Fig4}
\end{figure}
\par We also take down 25\% of nodes across multiple regions and measure latency in that case:
\begin{figure}[H]
    \includegraphics[width=\linewidth]{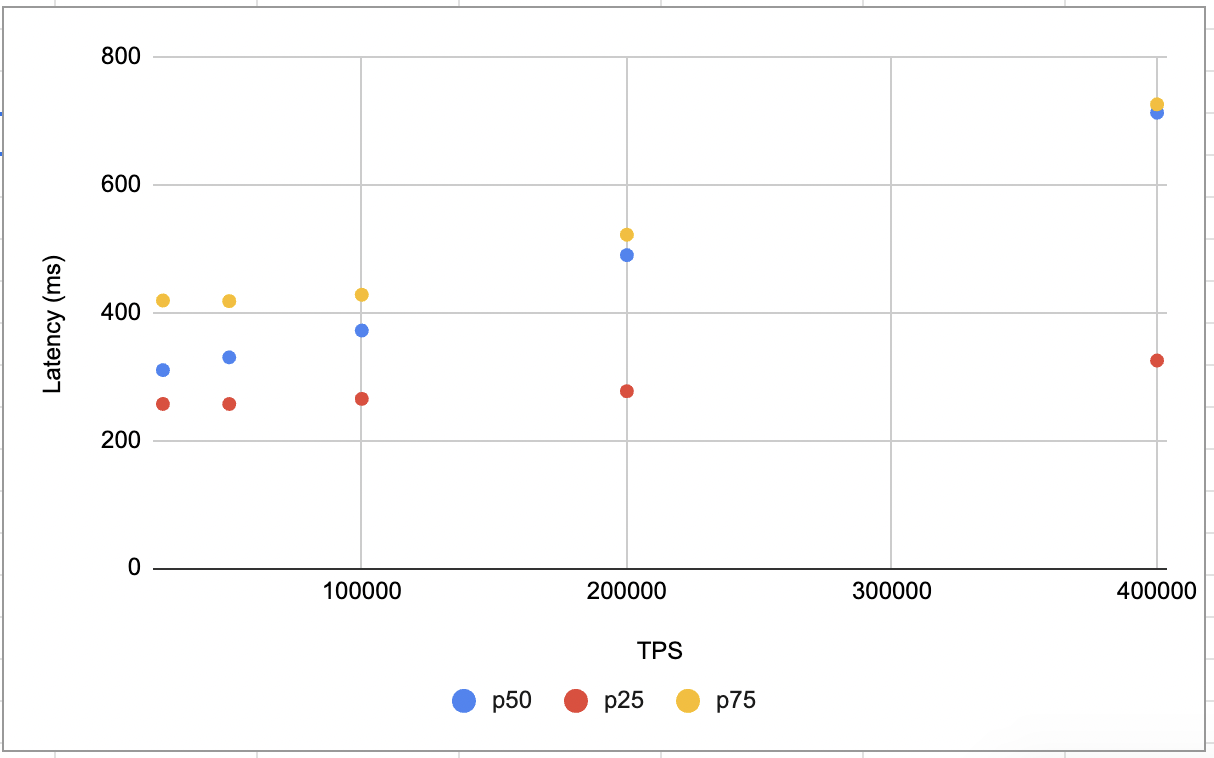}
    \caption{Latency with 25\% validators malfunctioning}
    \label{Fig5}
\end{figure}
\par As shown on figures, introducing failures does not affect latency significantly.
\par To assess reasons for growing latency we measure total CPU usage during the experiment as well as utilization of a critical core task that performs operations with the DAG that are hard to parallelize.
\par The latency growth happens primarily due to increased utilization of the main core loop, since small overhead is introduced when processing extremely large blocks.
\par The measured maximum TPS limit is achieved when overall CPU usages surpasses 85\% mark. After this point, the system loses stability and latency becomes hard to measure. While the latency is not predictable, the system is still however operational even at this load. The AWS instance type we used in the benchmark is a relatively low tier and using more powerful instances will likely push protocol even further in terms of maximum TPS.
\begin{figure}[H]
    \centering
    \includegraphics[width=1\linewidth]{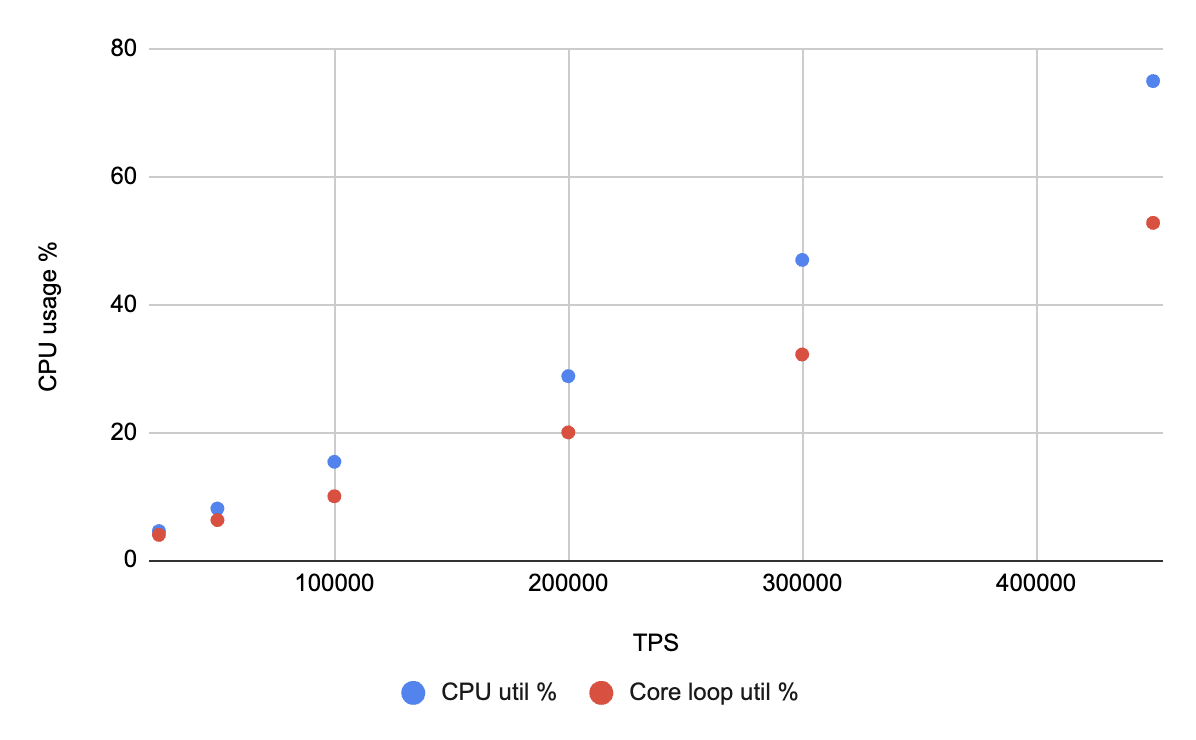}
    \caption{Cpu and core thread utilization}
    \label{Fig6}
\end{figure}
Note, that in the figure above, overall CPU utilization is counted as percentage of total CPUs used, while core thread CPU utilization is counted as percentage of a single CPU core.
\section{Conclusion}
\par In this paper, we have shown that DAG-based BFT consensus protocols can produce an outstanding performance, and when implemented correctly, we do not require to sacrifice latency for the high throughput. We also demonstrate that despite additional implementation challenges and extra steps required to protect DAG from Byzantine validators, an uncertified DAG approach comes with an undeniable performance benefits. As shown, it reduces the latency by at least 1 RTT comparing to the certified DAG protocols.
\par We also present bftd - performant Byzantine Fault Tolerance protocol implementation that can produce 450K TPS with 630 ms latency.
\section*{Acknowledgment}
We would like to thank George Danezis, Lefteris Kokoris-Kogias, Alberto Sonnino from the Mysten Labs for the feedback and for the great discussions that improved this work. We also extend our thanks to Kevin Nelson from the Aftermath Finance for the feedback that improved this work.
\bibliographystyle{plain}
\bibliography{references}
\end{multicols*}
\end{document}